\documentclass[slac_one]{revtex4}
\RequirePackage{lineno}
\usepackage{graphicx}
\usepackage{fancyhdr}
\begin{document}

\title{{\small{Hadron Collider Physics Symposium (HCP2008),
Galena, Illinois, USA}}\\ %% Please keep this conference title here
\vspace{12pt} Search for High-Mass Resonances Decaying into
Leptons of Different Flavor ($e\mu$, $e\tau$, $\mu\tau$)} %% Paper title goes here


\author{Yanjun Tu}
\affiliation{University of Pennsylvania, Philadelphia, PA 19104,
USA}

\begin{abstract}
We present a search for high-mass resonances decaying into two
leptons of different flavor: $e\mu$, $e\tau$ and $\mu\tau$. These
resonances are predicted by several models for Physics Beyond the
Standard Model, such as R-parity-violating MSSM. The search is
based on 1~${\rm fb^{-1}}$ of Tevatron Run II data collected with
the CDF detector at $\sqrt{s}$ = $1.96$ TeV in proton anti-proton
collisions.
\end{abstract}

\maketitle

\thispagestyle{fancy}


\section{Introduction} % Section title should be in all capitals.
One of the most promising theories for Physics Beyond the Standard
Model is Supersymmetry. In the minimal supersymmetric exstension
of the SM, the MSSM, an additional quantum number, defined as
\begin{eqnarray}
P_{R}=(-1)^{3(B-L)+2s}
\end{eqnarray}
is introduced. SM particles carry  R-parity of 1 while
supersymmetric particles have R-parity of -1. R-parity violating
(RPV) decays can happen through the following interaction terms:
\begin{eqnarray}
 W_{\Delta \rm L =1}& = & \frac{1}{2} \lambda^{ijk} L_i L_j \bar{e}_k +  \lambda^{'ijk} L_i Q_j \bar{d}_k + \mu^{'i} L_i  H_u , \nonumber \\
 W_{\Delta \rm B =1}& = & \frac{1}{2} \lambda^{''ijk} \bar{u}_i \bar{d}_j \bar{d}_k
\label{eq:W_Rp_odd}
\end{eqnarray}
where the indices i, j and k denote the generations. The fields in
Eq.~\ref{eq:W_Rp_odd} are superfields~\cite{Martin:1997ns}. The 48
RPV couplings in Eq.~\ref{eq:W_Rp_odd} are constrained from
theoretical and phenomenological points of
view~(\cite{Chemtob:2004xr},~\cite{Ledroit}).

The $\rm D\O$ Collaboration searched for heavy sneutrino decaying
into $e \mu$ in 1.0 ${\rm fb^{-1}}$ of data~\cite{Abazov:2007zz}.
The previous search for high-mass resonances decaying into $e \mu$
channel by the CDF Collaboration used 344~${\rm
pb^{-1}}$\cite{Abulencia:2006xm}.

In SUSY models, with an extra $U(1)'$, lepton flavor violation
(LFV) interactions are allowed by the Lagrangian density: $\rm
\frac{g_z'}{\sin \theta_{W}}[\bar{\psi_i}Q^{\psi}_{ij}\gamma^{\mu}
\psi_j]X_{\mu}$, where $i,j$ are generation indices, $g_Z'$ is the
$U(1)'$ gauge coupling, $\sin \theta_{W}$ is electroweak mixing
angle and $Q^{\psi}_{ij}$ is referred to as the
``charges''~\cite{Langacker:2000ju}. The lepton charges
$Q^{l}_{ij}$ are constrained by low energy
experiments~\cite{Murakami:2001cs}. CDF Run II performed a search
for $p \bar{p} \rightarrow Z' \rightarrow e\mu$ and excluded a
portion of the $Q^{l}_{12}$ vs.$M_{Z'}$
plane~\cite{Abulencia:2006xm}.

\section{Event Reconstruction}
\begin{figure}
\begin{center}
\scalebox{0.65}{\includegraphics[width=0.5\textwidth]{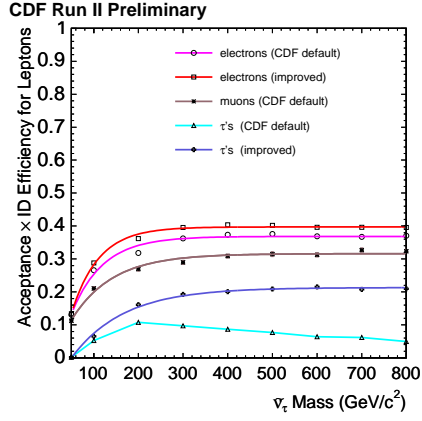}}
\end{center}

\caption{$\tau$ ID efficiency decreases when energy of $\tau$
increases if we use CDF default ID (light blue). The ID efficiency
of the $\tau$ keeps constant in the high energy region by using
the new ID (dark blue). In addition, $\tau$ ID efficiency has been
improved by factor of 2 by using the new ID. The ID efficiency of
electron has been improved by $\sim10\%$ by using the new ID.}
\label{fig:leptonacc}
\end{figure}

We use CDF Run II data corresponding to an integrated luminosity
of 1 $\rm fb^{-1}$. We select events with two identified leptons
of different flavor and opposite electric charge. To distinguish
the flavor, we select only hadronic taus. The kinematic cuts of
the leptons are the following: electron $\rm E_{T}\geq ~20 \ GeV$,
muon $\rm P_{T} \geq ~20 \ GeV/c$ and tau visible $\rm E_{T} \geq
~ 25 \ GeV$. Candidate events in $e\mu$ channel and $e\tau$
channel are collected by the high$\rm -p_{T}$ single electron
trigger path, which requires an electron candidate with $\rm
E_{T}\geq ~18 \ GeV $ and with pseudorapidity ${\rm |\eta|}
\lesssim$ 1.1. Candidate events in $\mu \tau$ channel are
collected by the high$\rm -p_{T}$ single muon trigger path, which
requires a muon candidate with $\rm P_{T}\geq ~18 \ GeV/c$ and
pseudorapidity ${\rm |\eta|} \lesssim$ 1.0. Lepton candidates
above the kinematic threshold are required to pass lepton
identification cuts. In this analysis, we select muons which pass
through CMUP (${\rm |\eta|} \lesssim$ 0.6) and CMX ( ${\rm |\eta|}
\lesssim$ 1.0) muon detectors~\cite{cdftdr}. They are required to
pass the standard CDF muon ID cuts~\cite{muoid}. Electron and
hadronic tau reconstruction and identification algorithms have
been improved to increase the acceptance at large lepton energy. A
standalone calibration of the Central Showermax detector
(CES)~\cite{cdftdr} has been performed, providing a better
resolution of the shower energy. As a consequence, the ID
efficiency of electrons improved by $\sim10\%$
(Figure~\ref{fig:leptonacc}). It also improves the energy
measurement of $\pi^{0}$ contributing to the visible momentum of
the hadronic tau~\footnote{We reconstruct hadronic taus through
their decay products. Tracks from charged particles ($\pi^{\pm}$
and $K^{\pm}$) are reconstructed by the Central Outer Tracker
(COT); neutral pion ($\pi^{0}$) energies and positions are
measured by the Calorimeter and Central Showermax Detector (CES).
We measure $\pi^{0}$ energy using the calibrated CES energy
measurement.}.

\begin{figure}
\begin{center}
\scalebox{0.65}{\includegraphics[width=0.5\textwidth]{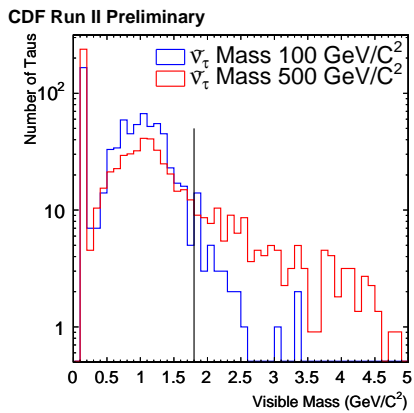}}
\end{center}

\caption[]{The visible mass distribution of $\tau '$s from 500
$\rm GeV/c^2$ ${\widetilde{\nu}}_{\tau}$ decay is broader than the
distribution of $\tau '$s from 100 $\rm GeV/c^2$
${\widetilde{\nu}}_{\tau}$ decay. The fixed mass cut ($\rm < \ 1.8
\ GeV/c^2$) is not efficient for high-energy $\tau '$s.}

\label{fig:vismass100500}
\end{figure}

\begin{figure}
\begin{center}
\scalebox{0.65}{\includegraphics[width=0.5\textwidth]{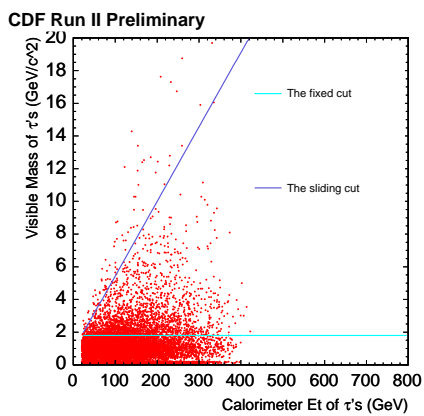}}
\end{center}

\caption[]{The sliding visible mass cut as a function of $\tau$
energy (dark blue line) and the fixed visible mass cut (light blue
line). The $\tau '$s between the dark blue line and light blue
line are recovered.}

\label{fig:slidingcut}
\end{figure}

As shown in Figure~\ref{fig:vismass100500}, the efficiency of the
standard fixed upper cut on the tau mass ($\rm < 1.8 \ GeV/c^{2}$)
decreases as the energy of the tau increases. The fixed cut on tau
visible mass (mass of the $\pi ^{0}$s and tracks originating in
the tau decay) is replaced by an energy dependent cut
(Figure~\ref{fig:slidingcut}). The upper cut is selected to yield
a constant 95\% efficiency when energy of the tau increases.

Tau quantities related to the hadronic shower are poorly simulated
and therefore tuned to data.

Due to the improvements described above, the ID efficiency of
$\tau$ improved by factor of 2 (Figure~\ref{fig:leptonacc}).

\section{Standard Model Backgrounds}
There are several SM processes having the same signature as the
signal: Drell Yan ($Z/\gamma^{*}\rightarrow\tau\tau$), diboson
($WW$) and $t\bar{t}$. There are also backgrounds due to
mis-identified leptons. We consider $W+$jet(s) and Drell
Yan$+$jet(s), where the gauge boson leptonically decays into
e/$\mu$/$\tau$ and a jet is mis-identified as a lepton. In
addition, leptons from Drell Yan ( $Z/\gamma^{*}\rightarrow ee$,
$Z/\gamma^{*}\rightarrow\mu\mu$) can be mis-identified; typically
the electron is mis-identified as tau and the muon is
mis-identified as electron or tau. These backgrounds are estimated
using PYTHIA Monte Carlo samples. The PYTHIA event generator with
CTEQ5L parton distribution function (PDF) and the CDF run II
detector simulation based on GEANT 3 are used to generate the
simulated samples.

 Along with the backgrounds listed above, QCD events with
jets faking either leptons, or with one jet faking a lepton (muon,
tau) and one $\gamma$ faking an electron are background to the
signal. To estimate this background, we use events with same
charge in data since there is no correlation between the charge of
the leptons in dijet and $\gamma+$jet events. We need to subtract
the contribution of same charge events originating from the
backgrounds other than dijets and $\gamma$ + jets.

\begin{figure}
\begin{center}
\scalebox{0.65}{\includegraphics[width=0.5\textwidth]{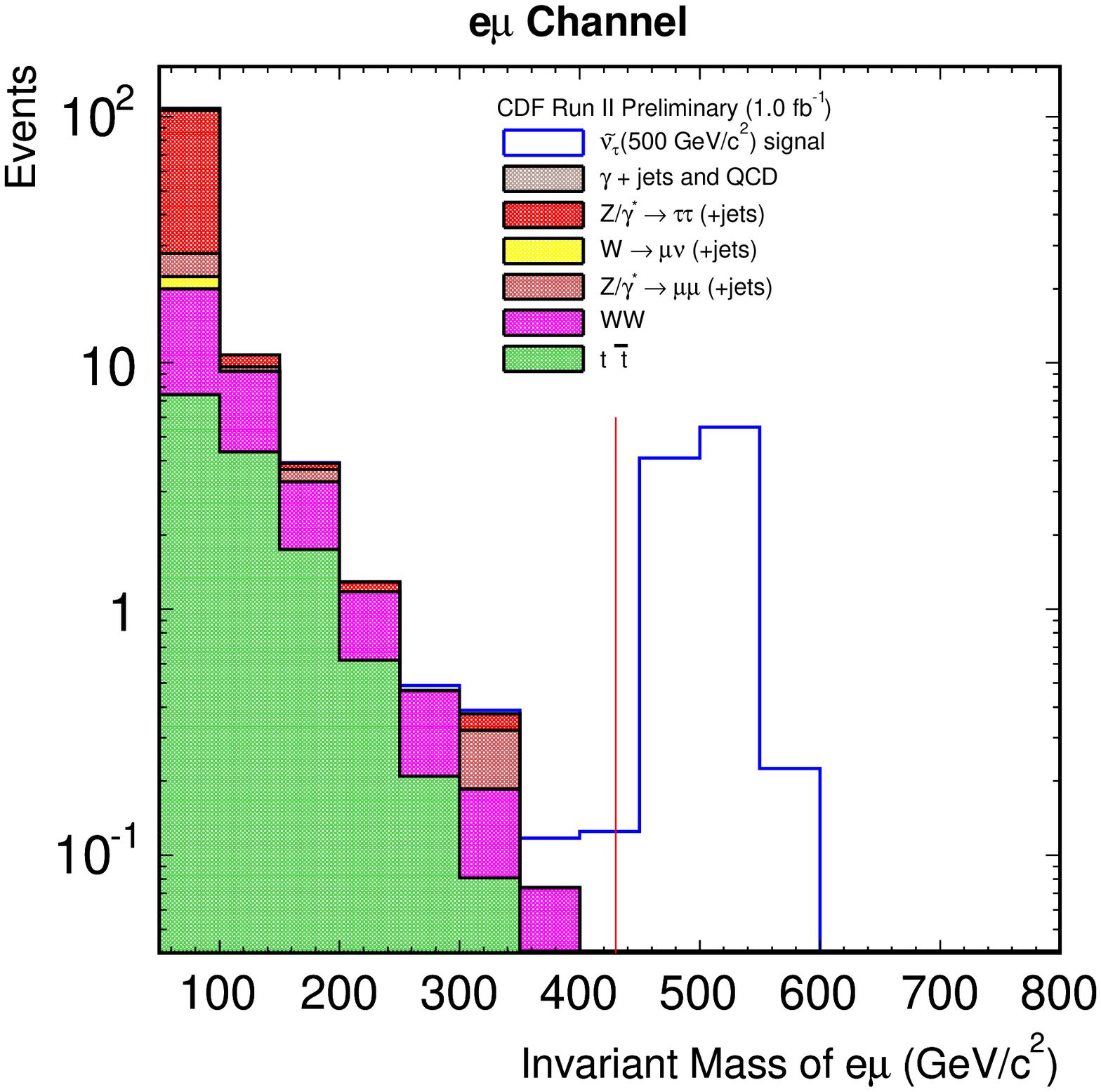}}
\scalebox{0.65}{\includegraphics[width=0.5\textwidth]{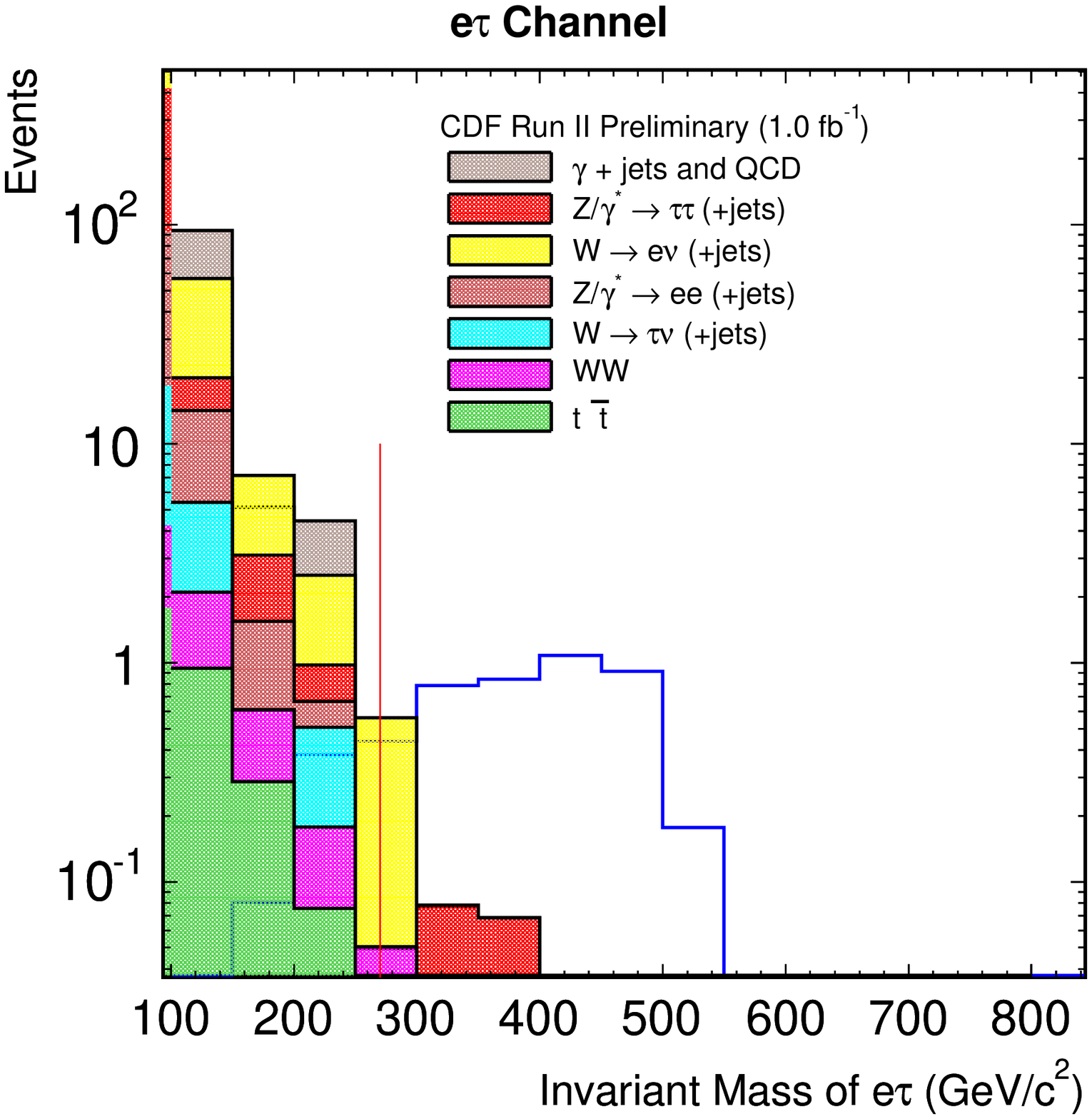}}

\end{center}
\caption[]{Dilepton mass distributions for SM background and
expected signal in e$\mu$ and e$\tau$ channels.}\label{figemusig}
\end{figure}

The background estimates in whole mass region are shown in
Fig~\ref{figemusig}.

\section{Control regions}
\begin{figure}
\begin{center}
\scalebox{0.65}{\includegraphics[width=0.5\textwidth]{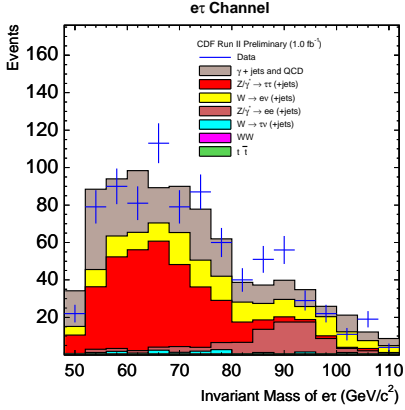}}
\end{center}
\caption[]{Dilepton (e$\tau$) mass in the e$\tau$ channel control
region.}\label{figcontrol_1}
\end{figure}

We check the consistency between the expectations from SM
backgrounds and data in the control region defined as: $\rm 50 \
GeV/c^{2} < dilepton \ mass \ M_{LL} <110 \ GeV/c^{2}$ (for
instance, Fig~\ref{figcontrol_1}). To estimate the number of
expected SM events, we use the same selection as applied to the
data. The events selected in the Monte Carlo background samples
are scaled by the lepton ID scale factors and the trigger
efficiencies.

\section{Conclusions}

We performed a search for high-mass resonances decaying into two
leptons of different flavor: $e\mu$, $e\tau$ and $\mu\tau$. The
search is based on 1~${\rm fb^{-1}}$ of Tevatron Run II data. We
first improved electron ID efficiency and tau ID efficiency. We
find agreement between the SM prediction and observation in data
in the control regions.

% figures should be put into the text as floats.
% Use the graphics or graphicx packages (distributed with LaTeX2e)
% and the \includegraphics macro defined in those packages.
% See the LaTeX Graphics Companion by Michel Goosens, Sebastian Rahtz,
% entire page. There is no need to do explicit centering.

% \begin{figure}
% \end{figure}

% Surround figure environment with turnpage environment for landscape
% figure
% \begin{figure}
% \end{figure}

%\begin{figure*}[t]
%\centering
%\includegraphics[width=135mm]{JACpic2.eps}
%\caption{Example of full width figure.} \label{JACpic2-f1}
%\end{figure*}

% If in two-column mode, this environment will change to single-column
% format so that long equations can be displayed. Use
%\begin{widetext}
%\end{widetext}

\begin{acknowledgments}
We thank the organizer of 2008 HCP conference for providing us a
chance to show the analysis. We also thank the CDF Collaborations
for their hard work which makes the results possible.
\end{acknowledgments}

%\begin{thebibliography}{9}   % Use for  1-9  references

\end{document}